\documentclass[pdflatex,sn-nature]{sn-jnl}
\usepackage{graphicx,xcolor}
\usepackage{amsmath,amssymb,amsfonts}
\usepackage{booktabs,longtable,array,multirow}
\usepackage{textcomp}
\usepackage{needspace}
\usepackage{placeins}
\usepackage{url}
\usepackage{xcolor}
\definecolor{re1}{RGB}{0,0,0}

\setcitestyle{super,sort&compress}
\hypersetup{hidelinks,bookmarksnumbered=false,pdftitle={Graph diffusion transformer for interpretable generation and inverse design of disordered atomic structures},pdfauthor={Jiani Hu; Chenyan Wang; Yongkun Yang; Ji Chen}}
\newcounter{suppnote}

\newif\ifshowrevision
\showrevisionfalse

\begin{document}
\raggedbottom
\setcitestyle{super,open={},close={}}
\title{Graph diffusion transformer for interpretable generation and inverse design of disordered atomic structures}
\author[1]{\fnm{Jiani} \sur{Hu}}
\equalcont{These authors contributed equally to this work.}
\author[1]{\fnm{Chenyan} \sur{Wang}}
\equalcont{These authors contributed equally to this work.}
\author[1]{\fnm{Yongkun} \sur{Yang}}
\equalcont{These authors contributed equally to this work.}
\author[1]{\fnm{Guanyu} \sur{Hou}}
\author[1]{\fnm{Jihaochen} \sur{Zhang}}
\author[2,3]{\fnm{Lei} \sur{Liu}}
\author*[1,3,4]{\fnm{Ji} \sur{Chen}}\email{ji.chen@pku.edu.cn}
\affil[1]{\orgdiv{School of Physics}, \orgname{Peking University}, \orgaddress{\city{Beijing}, \postcode{100871}, \country{China}}}
\affil[2]{\orgdiv{School of Materials Science and Engineering}, \orgname{Peking University}, \orgaddress{\city{Beijing}, \postcode{100871}, \country{China}}}
\affil[3]{\orgdiv{Interdisciplinary Institute of Light-Element Quantum Materials, Research Center for Light-Element Advanced Materials}, \orgname{Peking University}, \orgaddress{\city{Beijing}, \postcode{100871}, \country{China}}}
\affil[4]{\orgdiv{State Key Laboratory of Artificial Microstructure and Mesoscopic Physics, Frontiers Science Center for Nano-Optoelectronics}, \orgname{Peking University}, \orgaddress{\city{Beijing}, \postcode{100871}, \country{China}}}

\abstract{
Atomic disorder offers a potent degree of freedom to tailor material properties, yet the resulting vast configuration space obscures structure--property relations and hinders rational inverse design.
Here we introduce Graph Amorphous Diffusion Transformer (GaDiT), a generative model that combines dynamic periodic graphs, masked conditioning and an interval-weighted training-noise distribution.
GaDiT generates high-quality disordered structures while accelerating sampling by two orders of magnitude compared with existing corrector-based approaches, as demonstrated across two-dimensional amorphous carbon and three-dimensional amorphous silicon and silica.
Moreover, GaDiT supports conditional generation under single or combined structural features, physical properties and preparation conditions in a unified model, and enables inverse design that expands accessible structure--property ranges beyond the training distribution.
To interpret these structures and the trained network, we propose a conditioned atomic response mapping analysis, revealing both the global structural descriptors and the local atomic environments that govern target properties.
Overall, GaDiT establishes a general framework for interpretable generation, inverse design and structure--property discovery for disordered atomic structures.}

\maketitle
 
\phantomsection\label{sec-introduction}

Disordered materials have broad applications, ranging from common glass in daily life{\cite{IJAGS_2013_mauro2013glass}} to dielectrics in high-performance microchips\cite{Nature_2020_hong2020ultralow} and nonvolatile phase-change memory devices\cite{NM_2007_wuttig2007phase}. 
Because atomic disorder strongly affects electronic transport and mechanical response,\cite{PR_1958_anderson1958absence,JSP_2006_lemaitre2006sum,Nature_2023_tian2023disorder} 
a central challenge in disordered systems is establishing structure--property relations, namely, 
understanding how atomic configurations affect material properties.\cite{NRM_2025_liu2025amorphous}
Experimental studies of these relations face two main difficulties. 
First, atomic structures in disordered materials vary with preparation history\cite{NC_2018_luo2018ultrastable} and are often difficult to resolve at the atomic scale \cite{ChemSci_2022_elmachachi2022exploring,Nature_2021_yang2021determining}. 
Second, structural changes can affect material properties in complex ways across these diverse configurations, making it difficult to isolate the structural features responsible for a given response{\cite{PNAS_2014_ding2014soft,NP_2020_bapst2020unveiling}}.

Computational simulations are therefore essential. 
A common workflow involves two steps: first, disordered structures are generated using molecular dynamics (MD)\cite{JPCL_2018_deringer2018realistic,PR_1967_verlet1967computer} or Monte Carlo simulations\cite{PRL_1985_wooten1985computer,JPCL_2025_hu2025simulated}, and the properties of interest are calculated for these structures. Second, structural descriptors or order parameters are proposed to explain property variations across the generated structures\cite{PRB_2024_zhang2024structural,PRB_2013_bartok2013representing,JPCL_2025_gastellu2025disentangling,ChemRev_2021_musil2021physics}, with machine learning playing an increasingly important role in this step recently\cite{NP_2016_schoenholz2016structural,NP_2020_bapst2020unveiling}. However, the reliability and generality of the resulting relations depend on how well the simulation-generated dataset represents the relevant configuration space. 
Although independent simulations can be initialized from different configurations, conventional MD and local-move Monte Carlo still explore configuration space through sequential updates within each run. Transitions across energy barriers separating metastable configurations can be rare, particularly when they require collective atomic rearrangements or changes in network topology.\cite{PRX_2017_ninarello2017models,PRL_1985_wooten1985computer} 
Moreover, because practical MD trajectories are restricted to picosecond-to-nanosecond timescales, simulated melt--quench cooling rates are orders of magnitude faster than experimental synthesis. 
In amorphous silicon, silica and monolayer carbon, such rapid quenching traps networks in high-energy states with excess coordination defects\cite{JPCL_2018_deringer2018realistic,npjCM_2026_yang2026generative} or confines them to narrow structure--property windows that miss the diverse pore and domain morphologies observed experimentally\cite{Nature_2020_toh2020synthesis,Nature_2023_tian2023disorder}, leaving crucial regions of configuration space undersampled within practical computational budgets.
We therefore need methods that can efficiently sample a broader configuration space and directly generate structures with target features and properties.

Generative models provide a route to tackle these challenges by learning the distributions of atomic configurations and sampling structures under prescribed conditions.\cite{NeurIPS_2020_ho2020denoising,ICLR_2021_song2021score,NCS_2023_weiss2023guided,PRL_2025_li2025deep,NM_2026_cheng2026artificial} 
Property-directed inorganic crystal design demonstrates that target properties can guide structural sampling.\cite{Nature_2025_zeni2025generative,SB_2025_cao2025space,NCS_2026_cheng2026enhancing} Unified generation of molecules and periodic materials further shows that a shared generative representation can accommodate different bonding environments and boundary conditions,\cite{ICML_2025_joshi2025all} while diffusion-based biomolecular structure prediction uses sequence and molecular context to recover complex atomic arrangements.\cite{Nature_2024_abramson2024accurate}
However, extending this approach to disordered materials is more difficult.
Unlike crystalline systems, amorphous atomic networks lack long-range periodicity, and even single-element systems can exhibit diverse local environments and bonding topologies\cite{JACS_1932_zachariasen1932atomic}.
Unlike an isolated molecule, a periodic amorphous configuration represents a finite portion of an extended material.
Representative modelling therefore requires sufficiently large cells to capture structural heterogeneity and network connectivity, together with adequate sampling of distinct configurations to characterize material properties.\cite{npjCM_2026_yang2026generative,DD_2024_yong2024dismai}
Obtaining diverse reference configurations and their associated properties for such cells can be computationally demanding, limiting the availability of training data.\cite{AM_2026_finkler2026inverse,NCS_2024_benmahmoud2024data}
Recent studies have deployed diffusion and flow-based models built upon equivariant graph neural networks to generate amorphous silica and metallic glasses\cite{npjCM_2026_yang2026generative}, and to inverse-design multi-element glasses and amorphous oxides with targeted properties\cite{AM_2026_finkler2026inverse,SciAdv_2026_li2026boron,arXiv_2026_wu2026dmflow,arXiv_2026_cheng2026atlas}. 
{\color{re1}However, it remains difficult to achieve fast sampling, accurate energy distributions and flexible multi-property control at the same time\cite{DD_2024_yong2024dismai,AM_2026_finkler2026inverse}. Energy-based diffusion models obtain denoising forces by differentiating a learned scalar energy, which requires costly second-order derivatives during training and coordinate gradients at every sampling step\cite{AM_2026_finkler2026inverse,ICML_2025_bigi2025dark}. To suppress high-energy defects, existing diffusion workflows sometimes also add stochastic corrector steps such as Hamiltonian Monte Carlo, introducing thousands of extra evaluations per structure\cite{AM_2026_finkler2026inverse}. In addition, most conditional models assume a fixed set of input labels, and cannot generate structures under different combinations of targets without retraining.}

Here we introduce Graph Amorphous Diffusion Transformer (GaDiT), which combines a dynamic periodic graph transformer with masked conditioning and an interval-weighted noise distribution for generating disordered atomic structures. 
{\color{re1}Multi-head attention on dynamically updated periodic graphs captures short- and medium-range correlations. GaDiT predicts atomic noise vectors directly rather than differentiating a scalar energy, avoiding second-order derivatives during training and coordinate gradients during sampling. Concentrating training noise at intermediate amplitudes further sharpens property-targeting accuracy without extra sampling steps. 
Meanwhile, masked conditioning jointly encodes structural features, physical properties and preparation conditions while randomly hiding label subsets during training, allowing one unified model to support unconditional, single-condition and joint-condition generation. 
We validate GaDiT across three-dimensional amorphous silicon and silica as well as two-dimensional amorphous monolayer carbon (AMC)\cite{Nature_2020_toh2020synthesis}. In AMC, property-labelled training improves even unconditional generation, and joint conditioning directly generates structures with property combinations absent from the training set.
Structural analysis of these generated networks shows that global spatial heterogeneity, measured by ring-area inequality or atomic density, accounts for elastic-modulus variations beyond local crystalline order.
We further introduce Conditioned Atomic Response Mapping Analysis (CARMA), which tracks how internal atomic representations respond to changed target conditions, linking the model's learned modulus sensitivity to local structural distortions near hole boundaries and confirming the critical role of intermediate noise levels.
Through conditional generation, structural analysis and CARMA, GaDiT connects {\color{re1}fast and accurate} inverse design with physical interpretability, uncovering structure--property relations obscured in conventional simulations or missed by existing descriptors.}

\section{Results}\label{sec-19}

\subsection{GaDiT architecture}\label{sec-20}
Figure 1 summarises GaDiT's masked conditioning, interval-weighted noise distribution and periodic graph-transformer denoiser, together with the generation and evaluation workflow. One shared denoiser supports different conditioning requests and controlled changes in those inputs also provide a means to examine learned structure–property information. Dataset preparation is described in Methods, \hyperref[sec-68]{``Dataset preparation''}.

\begin{figure}[!htbp]
\centering
\includegraphics[width=\textwidth,height=0.68\textheight,keepaspectratio]{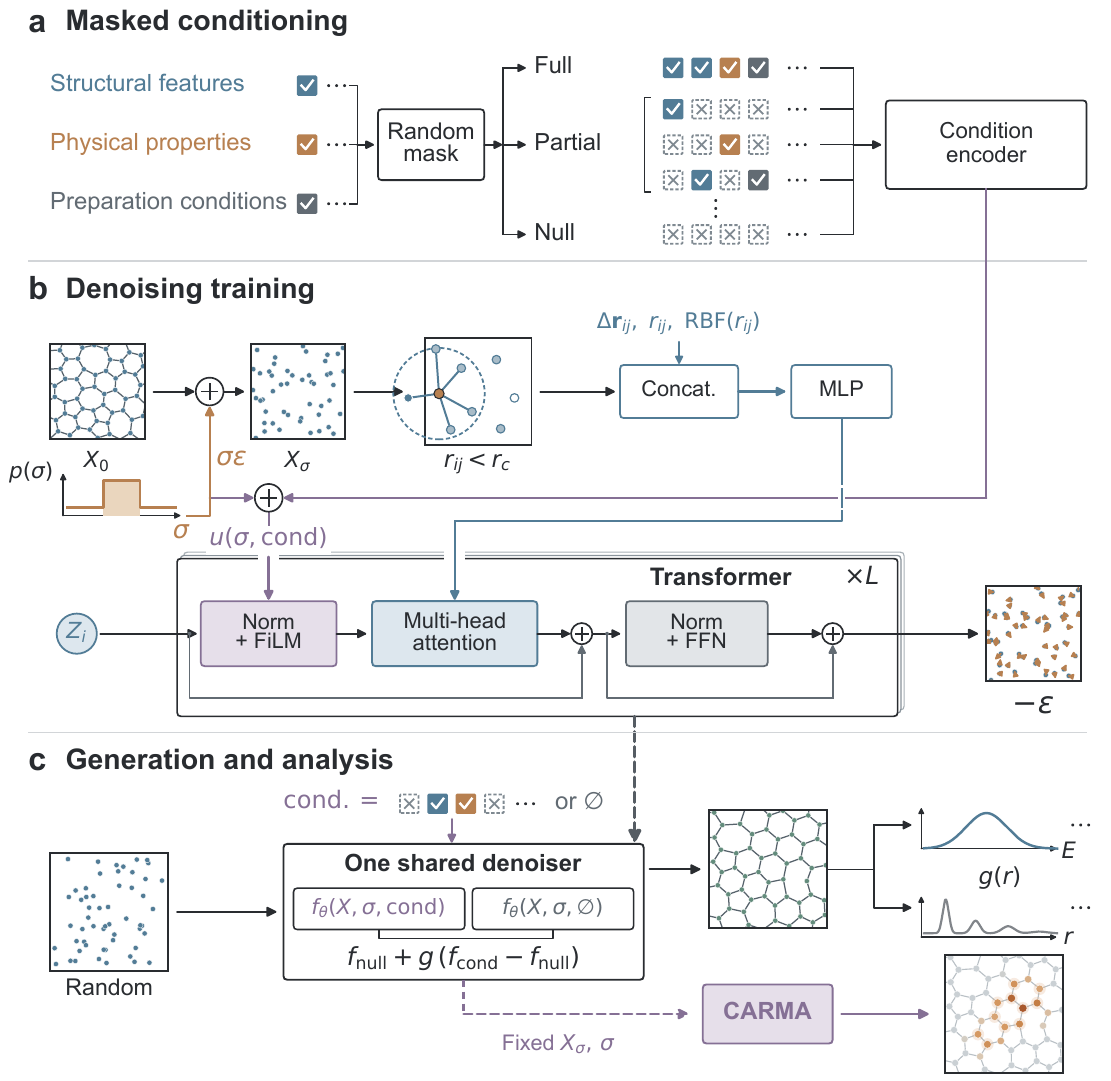}
\caption{GaDiT framework for generation of disordered atomic structures. 
a, Masked conditions. Training masks cover all subsets of structural features, physical properties and preparation conditions. 
Filled boxes with checkmarks and dashed boxes with crosses denote retained and masked labels, respectively.
b, Denoising training. Interval-weighted noise distribution determines coordinate perturbations. A Transformer operating on dynamic periodic graphs predicts the negative noise vector, $-\epsilon$, conditioned on the combined noise and condition embedding $u(\sigma,\mathrm{cond})$. $Z_i$ denotes element identity. Concat., concatenation; MLP, multilayer perceptron; Norm, layer normalization; FiLM, feature-wise linear modulation; FFN, feed-forward network. Circled plus symbols ($\oplus$) denote residual additions.
c, Generation and analysis. Selected targets remain fixed during sampling, while unspecified labels are masked. Conditional and unconditional (all-masked)  predictions from the same denoiser are combined with guidance scale $g$. A random configuration initializes guided sampling with the trained denoiser, followed by structural and property evaluation (such as the modulus $E$ and radial distribution function $g(r)$). Conditioned Atomic Response Mapping Analysis (CARMA) varies property conditions at fixed noisy coordinates, noise amplitude and trained parameters. Warm-coloured atoms highlight stronger schematic internal-feature responses.  
The dashed connection indicates reuse of the trained denoiser. 
}
\label{fig-1}
\end{figure}

To enable generation under different combinations of conditions without retraining, we train the denoiser with all (full-mode), some (partial-mode) or none (null-mode) of the available conditions, explicitly distinguishing masked entries from numerical targets (Fig.~\ref{fig-1}a). Structural features, physical properties and preparation conditions are supplied with a binary mask indicating which labels are retained. 
Masking operates on individual labels, and the sampling scheme includes every partial subset, as well as full and null inputs. Therefore, one shared model supports all $2^{K}$ conditioning subsets for $K$ labels, without a separate model for each subset. All masking patterns update shared denoiser parameters, so training with observed conditions can also shape the representation used in null-mode. 
The condition encoder and masking probabilities are specified in Methods, \hyperref[sec-70]{``GaDiT architecture and training''}.

The denoiser is a graph transformer drawing on transformer-based approaches to image generation,\cite{ICCV_2023_peebles2023scalable} molecular and crystal structure generation\cite{ICML_2025_joshi2025all,SB_2025_cao2025space} and protein structure prediction.\cite{Nature_2024_abramson2024accurate} GaDiT uses graph-dependent attention to aggregate information from neighbouring atoms on dynamically updated periodic graphs (Fig.~\ref{fig-1}b). Stacking attention blocks allows information to propagate across successive atomic neighbourhoods, offering a route to learning short- and medium-range structural correlations in amorphous materials. At each network evaluation, neighbours are reconstructed from the current coordinates using minimum-image periodic displacements. Edge features encode relative coordinates and interatomic distances, while a smooth cutoff weights attention near the neighbourhood boundary, similar to unconstrained point-edge transformers in atomistic modelling\cite{NeurIPS_2023_pozdnyakov2023smooth,NC_2025_mazitov2025pet}. Noise and condition embeddings modulate the transformer blocks. Architectural details are given in Methods, \hyperref[sec-70]{``GaDiT architecture and training''}.

During training, we draw coordinate-noise amplitudes from an interval-weighted distribution, assigning a higher probability density to a prescribed interval while retaining coverage of the full noise range (Fig.~\ref{fig-1}b). This allocation builds on the use of non-uniform training-noise distributions in diffusion models.\cite{NeurIPS_2022_karras2022elucidating,arXiv_2024_hang2024improved} 
At each drawn amplitude, mean-centred Gaussian perturbations are added to the clean coordinates, and the network predicts the dimensionless negative-noise vector $-\epsilon$. The weighting applies to the distribution of training noise amplitudes and the per-example denoising loss remains an unweighted mean-squared error. 
Low-amplitude perturbations preserve much of the local geometry, whereas larger perturbations obscure more of the original configuration. 
{\color{re1}Focusing more training samples on intermediate amplitudes balances local flexibility with structural context, lowering property-targeting errors more effectively than either uniform noise sampling or loss reweighting alone (Supplementary Note~1 and Supplementary Table~1).}
The noise interval, sampling density ratio and training settings are specified in Methods,\hyperref[sec-70] {``GaDiT architecture and training''}. 


The generation protocol begins from uniformly random atomic coordinates within a prescribed periodic cell.
During generation, the shared network evaluates the same noisy coordinates with the requested subset of conditions and with all conditions masked. Classifier-free guidance combines these predictions to steer denoising towards the requested targets,\cite{arXiv_2022_ho2022classifier} whereas null-mode generation uses the all-masked prediction (Fig.~\ref{fig-1}c).
The predictor follows a decreasing-noise schedule and is followed by deterministic refinement at the terminal noise level, without additional stochastic corrector updates (Fig.~\ref{fig-1}c). 
The guidance and update equations are given in Methods, \hyperref[sec-76]{``Conditional generation''}.

Alongside the generative model GaDiT, we introduce CARMA, a framework for examining learned structure--property relations. CARMA varies the supplied property condition while keeping the noisy coordinates, noise amplitude and trained network parameters fixed within each comparison. It combines atom-resolved feature-response mapping and local geometric analysis with denoising-error comparisons across noise levels, to connect the model's property dependence with structural environments (Methods, \hyperref[sec-local-response]{``Conditioned atomic response mapping analysis''}).

\FloatBarrier
\subsection{Unconditional generation and validation}\label{sec-34}

\begin{figure}[!htbp]
\centering
\includegraphics[width=\textwidth,height=1.0\textheight,keepaspectratio]{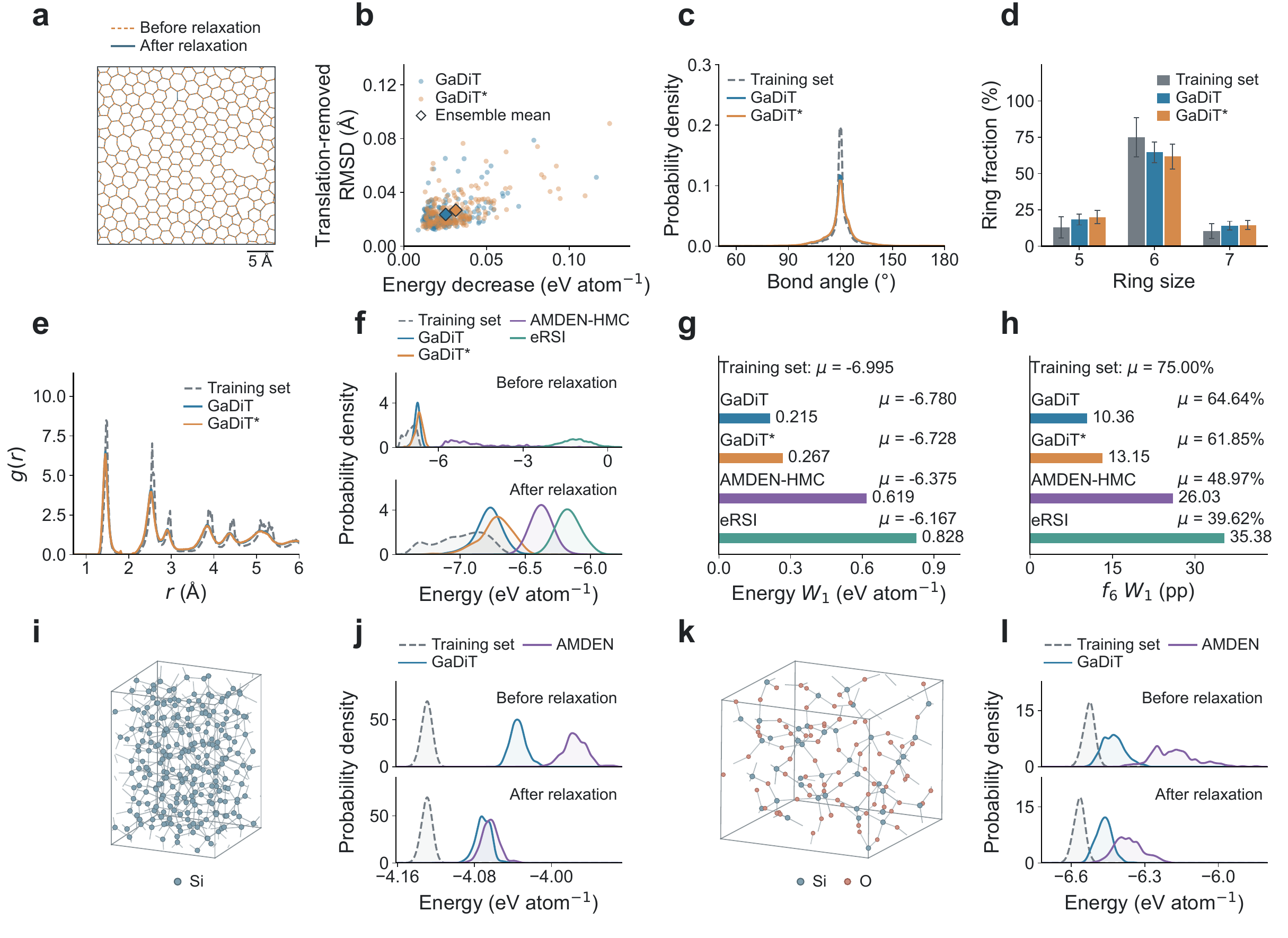}

\caption{Unconditional generation of two- and three-dimensional amorphous structures. a, Representative AMC configuration before and after fixed-cell relaxation. b, Paired energy decreases and translation-removed atomic root-mean-square displacements (RMSDs) after relaxation. Diamonds denote ensemble means. c--e, Bond-angle distributions (c), ring fractions (d) and radial distribution functions (e). Bars and error bars show means and sample s.d. f, AMC per-atom energy before and after relaxation. The full raw-energy range is shown in Supplementary Fig.~1. g,h, First Wasserstein distances ($W_1$) to the training set for relaxed energy and six-ring fraction $f_6$, with ensemble means ($\mu$) indicated. pp denotes percentage points. i,k, Representative relaxed GaDiT Si (i) and SiO$_2$ (k) configurations. j,l, Si (j) and SiO$_2$ (l) energy densities before and after relaxation. GaDiT denotes null-mode generation and GaDiT* property-free training; AMDEN and AMDEN-HMC omit and include the HMC corrector, respectively. Material-specific protocols are given in Supplementary Notes~2 and 3.}
\label{fig-2}
\end{figure}


We first evaluate generation without prescribed property targets, using AMC as the principal test system and extending the comparison to three-dimensional silicon and silica (Fig.~\ref{fig-2}). 
The AMC training set was prepared across a broad range of cooling rates to cover diverse degrees of structural order, whereas the silicon and silica tests use benchmark datasets from the AMDEN study\cite{AM_2026_finkler2026inverse} (Methods, \hyperref[sec-68]{``Dataset preparation''}).
GaDiT denotes the property-labelled model evaluated with all conditions masked and guidance scale $g=0$ (null mode).
To examine whether learning structure--property relations during training also benefits structure generation without prescribed targets, we compare GaDiT with a baseline model trained without property labels, denoted GaDiT*, as well as with existing amorphous structure generators\cite{AM_2026_finkler2026inverse,arXiv_2025_grenioux2025boltzmann}. 
Focusing first on AMC, we evaluate each model's raw outputs and configurations after identical fixed-cell relaxation with the Tersoff potential\cite{PRB_1989_tersoff1989modeling} separately (Supplementary Note~2).


The representative configuration and paired energy--displacement statistics indicate that the generated networks undergo only minor changes during fixed-cell relaxation (Fig.~\ref{fig-2}a,b). Across the 200 candidates generated by each model, the mean translation-removed atomic root-mean-square displacements (RMSDs) are 0.0236 and 0.0267~\AA{} for GaDiT in null mode and GaDiT*, respectively. The corresponding mean energy decreases are 0.0253 and 0.0314~eV~atom$^{-1}$. These small displacements and energy decreases indicate that the principal structural features are already established in the generated configurations before relaxation.

The bond-angle distributions (Fig.~\ref{fig-2}c) and radial distribution functions (Fig.~\ref{fig-2}e) reproduce the principal peak positions observed in the training set. Six-membered rings remain the most abundant ring type (Fig.~\ref{fig-2}d). The relaxed networks are predominantly threefold coordinated, with mean fractions of three-coordinated atoms of 95.40\%, 94.97\% and 98.05\% for GaDiT, GaDiT* and the training pool, respectively (Supplementary Fig.~2c). Predominantly threefold coordination, mixed ring sizes and broadened local geometry are also observed in atomic-resolution experiments on AMC.\cite{Nature_2020_toh2020synthesis} 
These results confirm that GaDiT generates realistic amorphous structures.

Having validated the physical plausibility of the generated networks, we next benchmark GaDiT against representative open-source amorphous generators, AMDEN with Hamiltonian Monte Carlo corrections (AMDEN-HMC)\cite{AM_2026_finkler2026inverse} and eRSI\cite{arXiv_2025_grenioux2025boltzmann}, to evaluate its advantages in distributional accuracy and sampling efficiency (Fig.~\ref{fig-2}f--h). Even without additional stochastic corrector updates, GaDiT produces energy and ring distributions substantially closer to the AMC training {\color{re1}set, while avoiding the high-energy tails of generated structures produced by AMDEN-HMC and eRSI (Supplementary Fig.~1).} 
After relaxation, the per-atom energy is $-6.780\pm0.096$~eV\,atom$^{-1}$ for GaDiT, compared with $-6.995\pm0.187$~eV\,atom$^{-1}$ for the training set (mean $\pm$ s.d.).
The generated energy distribution remains narrower than the deliberately broad training set.  
To quantify distributional agreement, we use the first Wasserstein distance $W_1$, which denotes the total area between the cumulative distribution curves of each generated ensemble and the training set (Supplementary Fig.~3). Smaller values indicate closer agreement.
{\color{re1}For both relaxed per-atom energy and six-membered-ring fraction ($f_6$), GaDiT reduces the $W_1$ errors to less than half of those for AMDEN-HMC and less than one-third of those for eRSI (Fig.~\ref{fig-2}g,h and Supplementary Table~2). Moreover, comparing null-mode GaDiT with GaDiT* highlights the benefit of property-labelled training, lowering both the relaxed-energy and $f_6$ $W_1$ errors by $\sim 20\%$ (from $0.2670$ to $0.2148$~eV\,atom$^{-1}$, and from $13.15$ to $10.36$ percentage points, respectively; Fig.~\ref{fig-2}g,h). Information learned from labelled examples can therefore also benefit generation with all labels masked, helping the shared network organize configurations across a broad range of structural disorder.}

{\color{re1}Beyond higher distributional accuracy, GaDiT achieves a favourable balance between structural fidelity and computational efficiency (Supplementary Note~2 and Supplementary Table~3). Because GaDiT directly predicts dimensionless noise vectors rather than differentiating a scalar pseudo-energy, training requires only standard first-order backpropagation instead of costly mixed second derivatives\cite{AM_2026_finkler2026inverse,ICML_2025_bigi2025dark}. On an identical GPU, null-mode GaDiT generates an AMC structure more than two orders of magnitude faster than energy-based AMDEN-HMC and also faster than a corrector-free score-based AMDEN variant\cite{AM_2026_finkler2026inverse}. Although the lightweight eRSI flow model runs at sub-second speed owing to its much smaller parameter count, its limited capacity sacrifices topological fidelity and yields nearly fourfold larger distributional errors.}

We further apply the framework to three-dimensional amorphous silicon and silica using datasets from the AMDEN study.\cite{AM_2026_finkler2026inverse} Representative relaxed networks generated by GaDiT are shown in Fig.~\ref{fig-2}i,k. For both materials, null-mode GaDiT produces energy distributions closer to the training reference than the tested AMDEN baseline, both before and after physical relaxation (Fig.~\ref{fig-2}j,l). 
{\color{re1}Using an order of magnitude fewer sampling steps than the silica baseline, GaDiT cuts the unrelaxed silica energy $W_1$ by nearly four-fifths (from $0.4486$ to $0.0996$~eV\,atom$^{-1}$) and halves the relaxed-energy error (from $0.2029$ to $0.1010$~eV\,atom$^{-1}$), while also lowering the relaxed-energy error for silicon (from $0.0661$ to $0.0589$~eV\,atom$^{-1}$; Supplementary Note~3).}
The silica GaDiT ensemble spans 144--249 atoms, demonstrating generation across different prescribed system sizes in a three-dimensional, multielement network. These tests extend the framework beyond planar, single-element AMC while maintaining closer energetic agreement without stochastic corrector updates. Material-specific training settings, initialization and relaxation protocols, three-dimensional structural checks are detailed in Supplementary Note~3.

\subsection{Conditional generation and inverse design}\label{sec-41}

\begin{figure}[!htbp]
\centering
\setlength{\belowcaptionskip}{2pt}
\includegraphics[width=\textwidth,height=0.67\textheight,keepaspectratio]{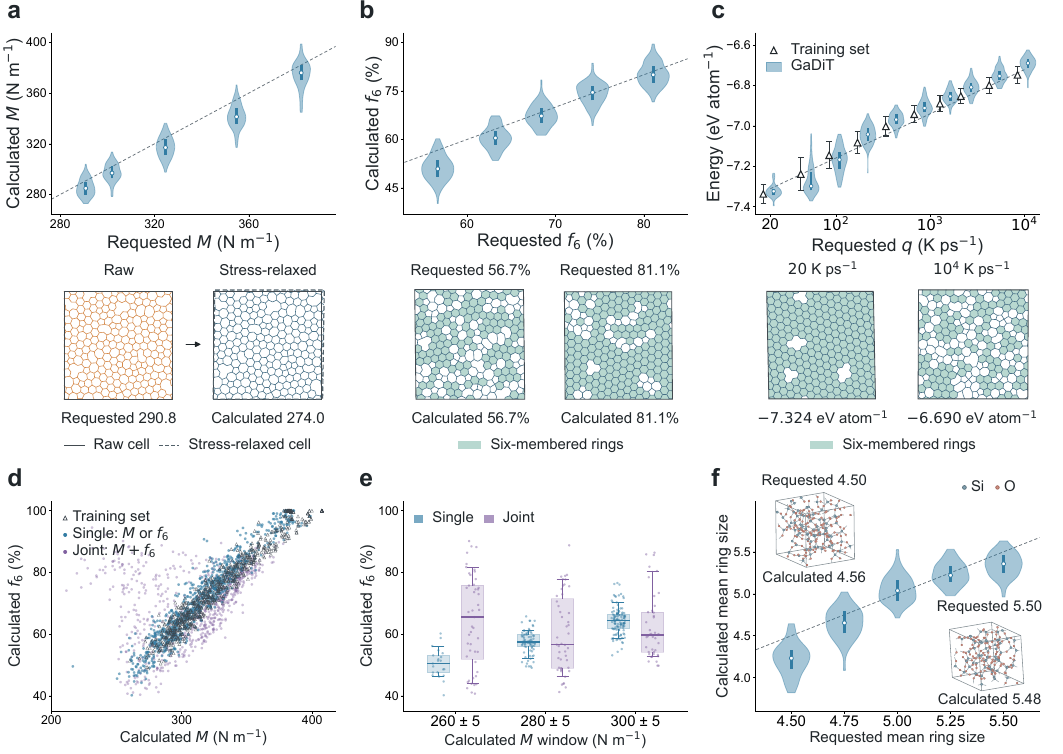}
\caption{Conditional generation across AMC and three-dimensional silica. a,b, Single-condition generation results for modulus $M$ (a) and six-membered-ring fraction $f_6$ (b). c, Per-atom potential energy after relaxation versus requested cooling rate $q$ across the training range ($20$--$10^4$~K\,ps$^{-1}$). Open triangles with error bars show training-set means $\pm$ s.d., and the dashed line shows their log-linear fit. Representative relaxed structures generated at $q=20$ and $10^4$~K\,ps$^{-1}$ are shown below. Violins show calculated distributions, in which white points and thick bars mark medians and interquartile ranges, and dashed lines indicate exact agreement between calculated and requested values. 
Below a, a representative generated structure before (raw) and after stress relaxation is shown at the same scale. Solid and dashed grey outlines denote their respective cells. Below b and c, representative structures at low, intermediate and high target $f_6$ (b) and at slow and fast requested $q$ (c) values are shown with six-membered rings shaded. d, Achieved $M$--$f_6$ pairs for the training set, single-condition generated structures ($M$ or $f_6$ alone) and joint-condition generated structures ($M + f_6$). e, Calculated $f_6$ within $M=260,280,300\pm5$~N\,m$^{-1}$. Points show all individual generated structures; boxes, centre lines and whiskers denote the 25th--75th percentiles, medians and 10th--90th percentiles. 
f, Calculated versus requested mean ring size (in Si atoms per ring) for stress-relaxed SiO$_2$ structures across five targets, with the dashed line indicating exact agreement. Insets show representative stress-relaxed structures near the group medians at $4.50$ and $5.50$. 
Complete protocols and material-specific ring definitions are given in Supplementary Note~4.}
\label{fig-3}
\end{figure}

We next examine single-condition generation across the three label categories in Fig.~\ref{fig-1}a, taking elastic modulus $M$, six-membered-ring fraction $f_6$, and cooling rate $q$ as representative targets for mechanical properties, network structure and preparation conditions, respectively. For each task, the requested condition is supplied while the remaining inputs are masked, and generation starts from independently drawn random coordinates (Fig.~\ref{fig-3}a--c). Linear fits to all individual outputs across five targets yield slopes of 0.948 and 1.194, with $R^2=0.928$ and 0.884 for $M$ and $f_6$, respectively. 
The calculated moduli of the generated structures closely match the requested targets with a near-unity slope, and both observables show nearly linear responses across the tested ranges. 
After relaxation, the individual-output targeting mean absolute error (MAE) is 9.31~N\,m$^{-1}$ for $M$ and 3.56 percentage points for $f_6$. 
{\color{re1}In controlled training ablations without corrector steps, interval-weighted noise sampling cuts these targeting errors by $40\%$ for $M$ and $32\%$ for $f_6$ compared with uniform noise sampling (Supplementary Table~1).} 
Because cooling rate $q$ is a dynamical preparation parameter rather than a static observable, we validate $q$-only generation against the energy--cooling-rate dependence of the melt--quench training set across $q = 20$--$10^4$~K\,ps$^{-1}$ (Fig.~\ref{fig-3}c). Generated energies closely follow the logarithmic cooling-rate trend of MD simulations ($R^2 = 0.991$) within the $20$--$10^4$~K\,ps$^{-1}$ training range. 

The structural examples illustrate how this control is expressed in the generated networks. Fig.~\ref{fig-3}a compares a raw output with its stress-relaxed reference, obtained by relaxing the atomic coordinates and in-plane cell before finite-strain modulus calculations. Displaying both states at the same physical scale makes the changes in cell shape apparent, and the reported modulus belongs to the relaxed reference. Fig.~\ref{fig-3}b shows two independently generated networks selected near the median calculated $f_6$ at low and high requests, while Fig.~\ref{fig-3}c contrasts a nearly crystalline network generated at slow cooling ($20$~K\,ps$^{-1}$), with a strongly disordered network generated at rapid quenching ($10^4$~K\,ps$^{-1}$). Larger requested fractions produce networks with more six-membered rings, as highlighted in the structural views. 
Sampling, mechanical evaluation and representative-selection criteria are detailed in Supplementary Note~4.

Beyond controlling individual properties, masked conditioning allows multiple conditions to be supplied simultaneously for multi-objective inverse design. We illustrate this capability by jointly conditioning on $M$ and $f_6$, combining mechanical and structural requirements within the same model. Fig.~\ref{fig-3}d compares the achieved $M$--$f_6$ distribution of structures generated under joint condition and single-condition requests. Joint-condition generation accesses property combinations absent from the training set, including low-modulus networks with large six-ring fractions. Within fixed modulus intervals, structures generated with joint conditions span a substantially broader range of $f_6$ than those generated under either single condition (Fig.~\ref{fig-3}e). At $M=280\pm5$~N\,m$^{-1}$, for example, the central 80\% of joint-condition generated structures spans $f_6=45.3$--$77.7$\%, compared with $51.5$--$61.6$\% for single-condition outputs. 
This ability to vary $f_6$ widely at a given modulus indicates that GaDiT captures underlying structure--property relations beyond the dominant trend in the training set, enabling the sampling of diverse network topologies within a target modulus range.
Target values, sample counts and additional conditioning analyses are provided in Supplementary Note~4.

Besides AMC, we also test how GaDiT performs in the conditional generation of three-dimensional disordered structures using the silica model introduced in Fig.~\ref{fig-2}. We use the mean ring size as the target generation property, which is defined as the mean number of Si atoms per ring, while masking the remaining labels. The insets in Fig.~\ref{fig-3}f show representative stress-relaxed SiO$_2$ structures generated at the lowest (4.50) and the highest (5.50) target ring sizes. 
Across five targets from 4.50 to 5.50 (50 generated structures per target, spanning 162--249 atoms), the post-relaxation ring-size distributions track the target values closely (Fig.~\ref{fig-3}f), with low MAEs of 0.210 and 0.173 Si atoms per ring before and after stress relaxation, respectively.
These results demonstrate effective conditional generation in a three-dimensional, multielement network across different prescribed system sizes. 
Pre-relaxation distribution and evaluation protocols are given in Supplementary Note~4.

\FloatBarrier
\subsection{Learning structure--property relations}\label{sec-49}

\begin{figure}[!htbp]
\centering
\includegraphics[width=\textwidth,height=0.65\textheight,keepaspectratio]{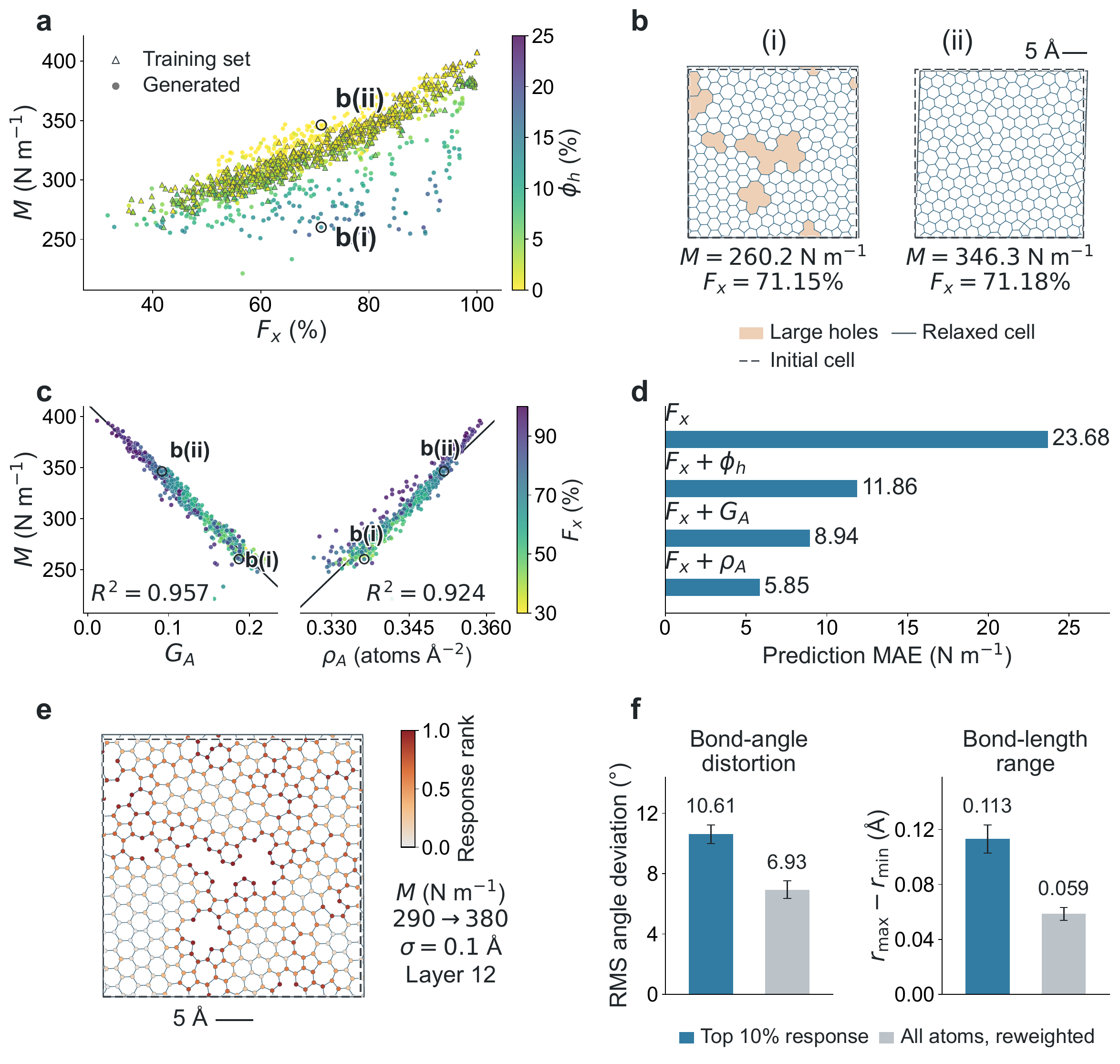}
\caption{Generated configurations and CARMA reveal complementary structure--property relations. 
a, Elastic modulus $M$ versus crystallite fraction $F_x$ across the joint-conditioned generated structures (circles) and the training configurations (triangles), coloured by large-hole area fraction $\phi_h$ (Supplementary Note~5). Outlined circles mark the two structures shown in b. 
b, Stress-relaxed structures with nearly identical $F_x$ ($71.15\%$ and $71.18\%$) but distinct moduli ($M=260.21$ and $346.26$~N\,m$^{-1}$). Large holes (rings with $\ge 10$ edges) are shaded, and solid and dashed outlines denote the relaxed and initial cells, respectively. 
c, $M$ versus ring-area Gini coefficient $G_A$ and stress-relaxed areal number density $\rho_A$ across the generated structures, colored by $F_x$, with linear fits and $R^2$ values indicated. 
d, Cross-validated MAE for estimating $M$ across the generated structures from $F_x$ alone or $F_x$ paired with $\phi_h$, $G_A$ or $\rho_A$ (evaluation protocols in Supplementary Note~5). 
e, Atom-resolved CARMA map for structure b(i), showing within-structure ranks of mean-centred layer-12 feature responses between target moduli of $290$ and $380$~N\,m$^{-1}$ at $\sigma=0.1$~\AA{} (averaged over two noise realizations). 
f, Bond-angle distortion (RMS deviation from $120^\circ$) and bond-length range ($r_{\max}-r_{\min}$) of the top 10\% most responsive atoms (blue) compared with a coordination-matched all-atom baseline (grey), averaged across $100$ generated structures on unperturbed relaxed coordinates ($d<1.85$~\AA{}; error bars denote 95\% bootstrap confidence intervals from 5000 structure resamples). All scale bars, $5$~\AA{}. 
}
\label{fig-4}
\end{figure}


In addition to inverse design, a trained generative model that captures how atomic arrangements govern material properties offers a powerful route to uncovering hidden structure--property relations. We pursue this goal from two complementary angles, first by generating structures in under-sampled regions of configuration space to identify missing physical descriptors, and second by probing the trained network itself to see which local atomic environments drive its property sensitivity. 
In Fig.~\ref{fig-3}, the structural target $f_6$ counts both isolated hexagons and crystalline domains, and thus reflects only short-range ring statistics. To capture medium-range crystalline order instead, we examine whether the crystallite fraction $F_x = N_{\mathrm{cryst}}/N_{\mathrm{rings}}$ proposed by Zhang et al.\cite{PRB_2024_zhang2024structural} is sufficient to explain elastic variation in AMC. Unlike $f_6$, $N_{\mathrm{cryst}}$ counts only six-membered rings in local crystalline clusters (three hexagons sharing a common vertex and adjacent edges) and $N_{\mathrm{rings}}$ is the total number of rings in the periodic cell (Methods). Their study showed that $F_x$ correlates closely with the elastic modulus $M$ in kinetic-Monte-Carlo-generated AMC, while noting that this relation could break down for configurations with markedly different domain or ring-area distributions that are rarely reached by standard sampling.\cite{PRB_2024_zhang2024structural} To access such configurations directly, we use GaDiT to perform joint $M$--$F_x$ generation across a $5\times5$ target grid (Supplementary Note~5). As shown in Fig.~\ref{fig-4}a, the generated structures expand well beyond the narrow $M$--$F_x$ correlation band of the melt--quench training set, producing configurations in which nearly identical $F_x$ values correspond to widely different elastic moduli.

To understand why structures with the same $F_x$ can differ so strongly in stiffness, we compare two representative generated configurations highlighted in Fig.~\ref{fig-4}a (shown in Fig.~\ref{fig-4}b). Despite sharing nearly identical crystallite fractions ($F_x=71.15\%$ and $71.18\%$), these two networks exhibit markedly different moduli of $260.21$ and $346.26$~N\,m$^{-1}$, respectively.
Visual inspection shows that the softer network in Fig.~\ref{fig-4}b(i) contains several large topological holes (shaded), whereas the stiffer network in Fig.~\ref{fig-4}b(ii) is uniformly compact. Because $F_x$ is a purely count-based ratio, each non-crystalline ring contributes only a single count to the denominator $N_{\mathrm{rings}}$ regardless of its physical size. Therefore a large void can soften the network substantially without noticeably changing $F_x$. To capture this spatial effect, we define the large-hole area fraction $\phi_h=\sum_{j\in\mathcal H}A_j/A_{\mathrm{cell}}$, where $\mathcal H$ denotes complete periodic rings with at least ten edges. 
Large holes account for $13.79\%$ of the relaxed cell area in the softer network of Fig.~\ref{fig-4}b(i), compared with $0\%$ in Fig.~\ref{fig-4}b(ii). Across the entire generated set in Fig.~\ref{fig-4}a, structures with lower moduli at a given $F_x$ consistently exhibit larger $\phi_h$ values, confirming that the area occupied by large holes is a key factor behind the modulus reduction.

Because $\phi_h$ only accounts for voids above a ten-edge threshold, we next ask whether broader spatial heterogeneity across all rings, even in the absence of large holes, also affects elastic stiffness. We examine two threshold-free geometric descriptors, the ring-area Gini coefficient $G_A$ and the relaxed areal number density $\rho_A=N/A_{\mathrm{cell}}$. 
Although all AMC structures are generated with $N=450$ atoms in an identical $36\times 36$~\AA$^2$ initial cell, $\rho_A$ varies across structures because the equilibrium cell area $A_{\mathrm{cell}}$ expands or contracts during variable-cell stress relaxation when calculating modulus (Fig.~\ref{fig-4}b).
For $n_f$ complete periodic rings with areas $A_i$, we define $G_A=\frac{\sum_{i=1}^{n_f}\sum_{j=1}^{n_f}|A_i-A_j|}{2n_f\sum_{i=1}^{n_f}A_i}$, which equals zero when all ring areas are identical and remains invariant under uniform rescaling. Whereas $\phi_h$ captures only large voids, $G_A$ quantifies relative ring-size inequality across the entire network and $\rho_A$ measures overall atomic density after stress relaxation.  Across the generated structures with defined ring partitions, $M$ decreases strongly with $G_A$ ($R^2=0.957$) and increases with $\rho_A$ ($R^2=0.924$; Fig.~\ref{fig-4}c). These strong correlations persist even when $F_x$ is restricted to a narrow interval ($67.5$--$72.5\%$) or among the generated structures without large holes ($\phi_h=0$, where $G_A$ retains $R^2=0.936$; Supplementary Note~5), confirming that ring-area inequality and areal density capture stiffness variations beyond both $F_x$ and large voids. 
This trend is consistent with previous AMC simulations showing that ring-network heterogeneity reduces elastic stiffness.\cite{EML_2020_ebrahem2020continuous} Physically, networks with more unequal ring sizes allow greater local bond-angle bending and non-affine atomic rearrangements under strain, which lowers the macroscopic modulus.\cite{JSP_2006_lemaitre2006sum} Topological ring-partition criteria and their relation to cell-area inequality statistics in two dimensional networks\cite{PRE_2024_klobusicky2024planar} are provided in Supplementary Note~5.


To compare the predictive power of these descriptors and to test whether accounting for spatial heterogeneity resolves the ambiguity of $F_x$, we fit empirical regression models that estimate $M$ from $F_x$ alone or from $F_x$ paired with each spatial descriptor (Fig.~\ref{fig-4}d; Supplementary Note~5). Using $F_x$ alone to estimate $M$ yields a large MAE of $23.68$~N\,m$^{-1}$, whereas pairing $F_x$ with $\phi_h$, $G_A$ or $\rho_A$ progressively reduces the prediction error to $11.86$, $8.94$ and $5.85$~N\,m$^{-1}$, respectively (Supplementary Note~5). 
Pairing local crystalline order ($F_x$) with a single global measure of spatial heterogeneity or atomic density thus cuts the modulus prediction error by up to fourfold. Crucially, none of $\phi_h$, $G_A$ or $\rho_A$ was supplied as a conditioning label during training. Instead, by generating controlled structures with matched $F_x$ but distinct moduli, GaDiT exposes the insufficiency of a purely count-based order parameter and guides the identification of these additional spatial features that account for modulus variations beyond the correlations sampled 
in conventional simulation datasets.

GaDiT captures these hidden structure--property relations without explicit spatial labels, 
raising the question of how the trained network itself perceives and utilizes modulus information at the atomic scale. To look inside the GaDiT network, we introduce CARMA, an analysis method analogous to a physical stimulus--response experiment on the frozen network. By holding the noisy atomic coordinates and model weights fixed while varying only the input modulus condition, CARMA directly reflects how the network responds to property changes. 
First, at the atomic level, we ask which local environments within a structure respond most strongly when the target modulus is changed. Feeding the same weakly perturbed structure ($\sigma=0.1$~\AA{}) into the frozen network under two target moduli ($M=290$ and $380$~N\,m$^{-1}$), we compute the change in each atom's internal feature vector between the two conditions (after subtracting the global mean shift across all atoms; Methods). Ranking atoms by this response magnitude reveals that the strongest modulus sensitivity localizes along irregular hole boundaries (Fig.~\ref{fig-4}e). To characterize the local bonding geometry of these high-response regions statistically, we compare the top 10\% most responsive atoms across $100$ generated structures against a coordination-matched baseline (Fig.~\ref{fig-4}f; Methods and Supplementary Note~6). These high-response atoms exhibit substantially greater bond-angle distortion ($10.61^\circ$ versus $6.93^\circ$) and larger bond-length variation ($0.113$~\AA{} versus $0.059$~\AA{}).
Physically, when the network is stretched, atoms at these distorted defect sites can easily bend their bond angles and shift their positions rather than directly stretching stiff covalent bonds\cite{JSP_2006_lemaitre2006sum,EML_2020_ebrahem2020continuous}. These regions therefore act as mechanical soft spots\cite{PNAS_2014_ding2014soft} that release local stress and lower the overall elastic modulus.
Notably, even though neither void statistics nor local bond distortions were supplied during training, the network autonomously learned to focus on these mechanically compliant defect environments when adjusting elastic stiffness.

Second, across different noise amplitudes $\sigma$, we use CARMA to identify which noise regime is most critical for coupling modulus information to the atomic structure.
We compare the network's denoising errors when supplied with a structure's true modulus or a modulus randomly reassigned from another structure, with all other labels masked (Supplementary Note~6). 
Even in a baseline model trained with uniform noise, the error reduction from supplying the true modulus peaks at intermediate amplitudes ($\sigma = 0.5$--$0.9$~\AA{}) and vanishes at low and high noise (Supplementary Fig.~13). This preference therefore arises naturally during denoising rather than from our noise-weighting scheme.
Because independent planar perturbations of amplitude $\sigma$ produce root-mean-square pairwise atomic displacements of $2\sigma$, this interval corresponds to relative displacements of $1.0$--$1.8$~\AA{}, spanning the $sp^2$ C--C bond length ($\sim1.4$~\AA{}). Without any prior knowledge of topological rules, the model autonomously learns that this bond-length scale is the critical regime governing elastic stiffness. 
When perturbations are much larger than the bond length (high $\sigma$), atomic positions are too disordered for precise reconstruction, whereas when perturbations are much smaller (low $\sigma$), the bonding topology is already settled and denoising reduces to local geometric refinement. However, when perturbations are comparable to the bond length, local neighbor connectivity remains flexible while the surrounding structural context is preserved, allowing the supplied modulus to guide topological reconstruction.
This finding directly accounts for the effectiveness of our interval-weighted noise distribution within this bond-length regime ($\sigma=0.5$--$0.9$~\AA{}), demonstrating that CARMA reveals both the local defect sites and the physical length scale of structure--property coupling learned by the model.

\FloatBarrier
\section{Discussion}\label{sec-61}
Rather than treating generative models solely as black-box tools for inverse design, this work establishes GaDiT and CARMA as an integrated framework for both generating disordered materials and uncovering their underlying structure--property relations.
Using one trained network without stochastic corrector steps, GaDiT generates physically realistic structures of two-dimensional AMC {\color{re1}(two orders of magnitude faster than Monte-Carlo-corrected diffusion)}, three-dimensional amorphous silicon and silica with variable system sizes. 
Notably, sharing representations across structural, mechanical and processing labels improves even unconditional generation, indicating that learning structure--property couplings helps the model capture the intrinsic organization of disordered networks. 
Unlike conventional MD or Monte Carlo simulations, joint conditioning directly samples structure--property combinations that are rare or absent in the training set. 
In AMC, this capability reveals that the crystallite fraction alone does not determine elastic stiffness. 
Global spatial heterogeneity instead accounts for the modulus variation, as captured by a single global measure such as ring-area inequality or atomic density.
Probing the frozen network with CARMA further shows that modulus conditioning focuses on distorted bonding environments along defect boundaries and acts mainly at bond-length-scale noise levels, which in turn provides a physical basis for interval-weighted noise training.

Beyond the elastic response of AMC, this controlled-generation and network-probing strategy can be applied to other physical properties and disordered systems. 
Comparing generated structures, which match in known descriptors but differ in a target property, can reveal additional structural features governing thermal transport\cite{PRB_1993_allen1993thermal}, electronic localization\cite{PR_1958_anderson1958absence,Nature_2023_tian2023disorder} or plastic deformation\cite{PNAS_2014_ding2014soft} in amorphous solids.\cite{NP_2020_bapst2020unveiling} Similarly, CARMA can 
reveal both the local atomic environments and the critical noise scales that govern these properties inside the trained network. Because GaDiT operates on dynamic periodic graphs and supports multiple elements and variable system sizes, it can be readily extended to other disordered materials. Promising targets include amorphous boron nitride monolayers\cite{Nature_2020_hong2020ultralow,PRB_2024_zhang2024structural}, multicomponent oxide and metallic glasses\cite{NC_2018_luo2018ultrastable,SciAdv_2026_li2026boron,AAAI_2026_chen2026physical}, phase-change chalcogenides\cite{NM_2007_wuttig2007phase} and disordered grain boundaries.

Our findings also offer practical insights for experimental studies of disordered materials. 
Because experimental synthesis such as chemical vapour deposition proceeds over much longer timescales than picosecond MD melt--quench simulations, real AMC samples in transmission electron microscopy\cite{Nature_2020_toh2020synthesis,Nature_2023_tian2023disorder} often contain voids and density variations, which lie outside idealized simulation trends but are directly accessible by joint-condition generation.
Our results suggest that experimental structure--property analysis should look beyond simple ring counts and measure ring-area inequality and local bond distortions around pores. For three-dimensional amorphous solids, resolving all atomic coordinates experimentally remains difficult.\cite{Nature_2021_yang2021determining} In future work, masked conditioning could be extended to take 
diffraction patterns, vibrational spectra, partial microscopy images or synthesis parameters as inputs\cite{MLST_2024_kwon2024spectroscopy,arXiv_2026_guo2026generative} to generate realistic atomic structures that match experimental measurements.

Several improvements could further expand the capabilities of this framework. First, jointly diffusing periodic cell vectors and atomic coordinates would allow direct constant-pressure generation and the study of pressure-induced phase transitions\cite{Nature_2021_deringer2021origins}. Second, coupling GaDiT with machine-learning interatomic potentials\cite{JPCL_2018_deringer2018realistic,ChemSci_2022_elmachachi2022exploring,NC_2025_mazitov2025pet} in an active-learning loop could iteratively add training data in sparsely sampled property regimes. Last but not least, incorporating free-energy estimates\cite{PRL_2025_li2025deep,arXiv_2026_li2026differentiable,arXiv_2026_cheng2026atlas} or growth-kinetics constraints into guided sampling would help 
evaluate the thermodynamic stability and experimental synthesizability of designed structures.
By bridging property-directed generation with mechanistic interpretability, GaDiT and CARMA provide a general paradigm for inverse materials design and physical discovery in disordered systems.

\section{Methods}\label{sec-67}

\subsection{Dataset preparation}\label{sec-68}

The source dataset comprises 1,000 planar configurations, each containing 450 carbon atoms in a 36 \ensuremath{\times} 36 \ensuremath{\times} 20 \AA$^{3}$ periodic cell. These structures were prepared by melt--quench molecular dynamics (MD)  in LAMMPS\cite{CPC_2022_thompson2022lammps} with the Tersoff potential\cite{PRB_1989_tersoff1989modeling,PRB_2010_lindsay2010optimized}. 
Starting from a pristine graphene supercell, carbon vacancies were introduced to yield 450 atoms and the in-plane cell size was slightly rescaled to $36 \times 36$~\AA$^{2}$. This geometry corresponds to an areal density of $0.347$~atoms\,\AA$^{-2}$, slightly below that of pristine graphene to accommodate amorphous ring rearrangements and pores. Each configuration was then heated to 12,000 K under NVT conditions, held for 50 ps, and cooled to 300 K at cooling rates spanning 20--10,000 ~K\,ps$^{-1}$ prior to final relaxation. Such a protocol can generate AMC structures ranging from highly disordered configurations to nearly crystalline structures containing defects. 
This broad cooling-rate range enables sampling across a wide range of structural disorder. Dataset filtering and training/validation/test partitioning are detailed in Supplementary Note~1.

To construct conditioning labels for each configuration and to evaluate generated structures consistently, we calculate the in-plane elastic modulus $M$ and the structural descriptors $f_6$ and $F_x$ as follows. 
For the mechanical label $M$, each configuration is first relaxed at fixed cell dimensions and then further relaxed by allowing the cell vectors to adjust until the in-plane stress vanishes.
Positive and negative normal and shear strains of amplitude 0.001 are then applied to this stress-free reference cell to obtain a symmetrized stiffness matrix, whose inverse is the compliance tensor $S$. The mean in-plane Young's modulus is defined as $M=(1/S_{11}+1/S_{22})/2$, reported in N~m$^{-1}$. Stress conversion and numerical convergence criteria are given in Supplementary Note~4.

For the structural labels $f_6$ and $F_x$, covalent bonds are defined between atoms with periodic distances shorter than 1.85~\AA. The six-membered ring fraction reads $f_6=N_6/\sum_{n=3}^{9}N_n$, where $N_n$ counts the number of unique chordless cycles of length $n$, excluding cycles winding around the cell. The crystallite fraction $F_x=N_{\mathrm{cryst}}/N_{\mathrm{rings}}$ quantifies the fraction of hexagons in mutually edge-adjacent triples among all complete periodic rings. This counting-based descriptor follows the crystallite criterion of Zhang et al.,\cite{PRB_2024_zhang2024structural} with the periodic construction specified in Supplementary Note~5.

The three-dimensional Si and silica calculations use separately trained models and datasets derived from the AMDEN source structures\cite{AM_2026_finkler2026inverse}. Physical relaxations were performed using the Stillinger--Weber potential\cite{PRB_1985_stillinger1985computer} with the FIRE algorithm\cite{PRL_2006_bitzek2006structural} for Si and the Munetoh Tersoff potential\cite{CMS_2007_munetoh2007interatomic} for silica. Their preparation, data splits, model-specific settings and sampling initializations are specified in Supplementary Note~3.

\subsection{GaDiT architecture and training}\label{sec-70}

GaDiT stacks modulated graph-transformer blocks, in which periodic neighbors are dynamically reconstructed at every evaluation within a predefined cutoff radius.
Cartesian and radial edge features enter smooth-cutoff-weighted attention. Noise and condition embeddings modulate each block. 
Writing $u=e_\sigma(\sigma)+e_{\mathrm{cond}}(\mathrm{cond.})$, the initial node features are $h_i^0=e_Z(Z_i)+u$. Each block applies layer normalization and feature-wise linear modulation (FiLM),\cite{AAAI_2018_perez2018film} followed by local multi-head attention with a residual update. A second normalization and feed-forward sublayer have a separate residual update.

To place different conditioning labels on a common numerical scale, each condition variable is standardized to zero mean and unit variance using the training-set mean and standard deviation, with the quench rate first log-transformed. Masked values are replaced with learnable placeholders (trainable embedding parameters representing withheld inputs). The resulting condition features are concatenated with a binary mask indicating which labels are supplied (1 for a supplied label and 0 for a withheld label), and the combined representation is encoded by a multilayer perceptron. Fully masked inputs use a separate learned null embedding.  
During training, conditioning inputs are randomly masked across full, null and partial regimes 
so that a single model learns the coupled structure--property dependencies and generates structures across all $2^K$ subsets of the $K$ labels. Masking probabilities are given in Supplementary Note~1.

For ground-truth coordinates $X$, training uses $X_\sigma=X+\sigma\epsilon$, where $\epsilon$ is mean-centered Gaussian noise. The denoiser $f_\theta$ predicts the dimensionless vector $-\epsilon$ with loss $\ell_\theta=\|f_\theta(X_\sigma,\sigma,\mathrm{cond.})+\epsilon\|_F^2/(3N)$, where $\mathrm{cond.}$ denotes the supplied conditioning values together with their mask, as illustrated in Fig.~\ref{fig-1}. Predicting $-\epsilon$ directly allows standard first-order backpropagation, avoiding the costly second-derivative calculations needed in energy-based diffusion models\cite{AM_2026_finkler2026inverse,ICML_2025_bigi2025dark}. 

An interval-weighted training-noise distribution increases the sampling density within a prescribed interval $[\sigma_{\mathrm{low}},\sigma_{\mathrm{high}}]$
while retaining coverage of the full noise range $[\sigma_{\min},\sigma_{\max}]$ 
using the piecewise-constant density 
\begin{equation}
p(\sigma)=\frac{1}{Z}
\begin{cases}
w_{\mathrm{in}}, & \sigma_{\mathrm{low}} \leq \sigma \leq \sigma_{\mathrm{high}},\\
1, & \sigma \in [\sigma_{\min},\sigma_{\mathrm{low}})\cup(\sigma_{\mathrm{high}},\sigma_{\max}],\\
0, & \text{otherwise},
\end{cases}
\end{equation}
where $w_{\mathrm{in}}>1$ is the in-band sampling-density ratio and $Z$ is the normalization constant. 
The weighting applies only to noise sampling and the per-example denoising loss remains an unweighted mean-squared error. 

Across all material systems, model parameters are optimized using AdamW
\cite{ICLR_2019_loshchilov2019decoupled} with gradient clipping. Network dimensions, noise-interval bounds, optimizer settings and material-specific configurations for AMC, amorphous silicon and silica are detailed in Supplementary Notes~1 and 3.

\subsection{Conditional generation}\label{sec-76}

Structure generation begins by randomly sampling fractional atomic coordinates from a uniform distribution within the prescribed periodic cell. Specifically, for two-dimensional amorphous monolayer carbon, coordinates are initialized in the plane while the out-of-plane coordinate is held fixed; for three-dimensional amorphous silicon and silica, all three spatial coordinates are sampled.
Sampling then proceeds through $N_{\mathrm{pred}}$ predictor steps along a decreasing noise schedule from $\sigma_{\max}$ to $\sigma_{\min}$, followed by $N_{\mathrm{ref}}$ refinement steps.

With $f_{\mathrm{cond}}$ and $f_{\mathrm{null}}$ denoting the dimensionless negative-noise predictions for requested and fully masked conditions at identical coordinates and noise amplitude, classifier-free guidance gives $f_{\mathrm{CFG}}=f_{\mathrm{null}}+g(f_{\mathrm{cond}}-f_{\mathrm{null}})$, where $g$ is the guidance scale.\cite{arXiv_2022_ho2022classifier}
The score used in coordinate updates is $s_{\mathrm{CFG}}=f_{\mathrm{CFG}}/\sigma$. Null-mode generation corresponds to $g=0$.

For structure generation, we use a score-based reverse-diffusion update.\cite{ICLR_2021_song2021score}  
Defining \(\Delta_i=\sigma_i^2-\sigma_{i+1}^2\), each step updates the coordinates as \(X\leftarrow X+\Delta_i s_{\mathrm{CFG}}+\sqrt{\Delta_i}\,\xi\), where \ensuremath{\xi} is zero-centered Gaussian noise (planar for AMC and three-dimensional for silicon and silica). 
After wrapping atomic positions back into the periodic cell, the periodic graph is reconstructed prior to each model evaluation. The $N_{\mathrm{ref}}$ deterministic refinement steps then apply \(X\leftarrow X+(\sigma_{\min}^2/N_{\mathrm{ref}})s_{\mathrm{CFG}}\) at fixed \(\sigma_{\min}\).\cite{ICLR_2021_song2021score}

All generated structures follow this predictor-and-refinement procedure without stochastic corrector updates. 
Material-specific sampling schedules, step counts and initialization cells for AMC, 
silicon and silica are documented in Supplementary Notes~1 and 3.

\subsection {Conditioned atomic response mapping analysis}\label{sec-local-response}
We developed Conditioned Atomic Response Mapping Analysis (CARMA) to probe how the trained GaDiT couples target properties to local atomic environments and noise amplitudes $\sigma$ within a unified analysis framework. Inspired by counterfactual condition-probing strategies in image diffusion models\cite{ICCV_2023_li2023your,ICLR_2023_couairon2023diffedit}, 
CARMA perturbs a relaxed structure $X$ with zero-mean Gaussian noise $\epsilon$ at amplitude $\sigma$ to form a noisy input $X_\sigma=X+\sigma\epsilon$. It then feeds this identical $X_\sigma$ into the frozen network under two distinct values of a single target condition $c$ ($c_1$ versus $c_2$), while keeping all other condition inputs masked. Because the atomic coordinates and network weights are identical in both evaluations, any difference in the network's denoising prediction or internal features is caused solely by the change in $c$. This design enables both atom-resolved and noise-resolved analyses.


In the atom-resolved analysis, CARMA evaluates the same perturbed structure $X_\sigma$ at a fixed low amplitude $\sigma$ under two prescribed target values $c_1$ and $c_2$. 
For transformer block $\ell$, the per-atom hidden-feature change and mean-centered response magnitude are defined as
\begin{align}
\Delta\mathbf h_i^{(\ell)}&=\mathbf h_i^{(\ell)}(X_\sigma,\sigma,c_2)-\mathbf h_i^{(\ell)}(X_\sigma,\sigma,c_1),\\
S_i^{(\ell)}&=\left\langle\left\|\Delta\mathbf h_i^{(\ell)}-\frac1N\sum_{j=1}^N\Delta\mathbf h_j^{(\ell)}\right\|_2\right\rangle_{\epsilon},
\end{align}
where subtracting the configuration-mean feature shift isolates spatial variations across the atomic network and $\langle\cdot\rangle_\epsilon$ averages over 
independent noise realizations $\epsilon$ applied to the same structure $X$ to suppress random-perturbation variance. Ranking atoms by $S_i^{(\ell)}$ identifies the local sites most sensitive to the target property. We then characterize the intrinsic bonding geometry of these top-ranking atoms in the original relaxed structure $X$ (before the probe noise $\epsilon$ was added). Because high-response sites often contain a higher proportion of under-coordinated atoms than the rest of the structure, directly comparing them against a simple average of all atoms would confound local geometric distortion with coordination-number differences. To rule out this confounding effect, we compare the selected top-ranking atoms against a coordination-matched baseline in the same relaxed structure $X$, constructed by weighting the per-coordination-number averages of $X$ to match the exact coordination composition of the selected atoms.

In the noise-resolved analysis, CARMA sets $c_1$ to the structure's true 
property label (matched) and $c_2$ to a label randomly reassigned from another configuration (shuffled) across a series of noise amplitudes $\sigma$. 
We define the matched-label advantage as $A(\sigma)=E_{\mathrm{shuffled}}(\sigma)-E_{\mathrm{matched}}(\sigma)$, where $E(\sigma)$ is the configuration-averaged mean-squared error in predicting the dimensionless negative noise $-\epsilon$. Positive $A(\sigma)$ indicates that supplying the true property label improves denoising accuracy at noise scale $\sigma$.

In this study, we apply CARMA to the elastic modulus $M$ with $q$, $f_6$ and $F_x$ masked. 
For the atom-resolved probe, we evaluate transformer block $\ell=12$ between $M_1=290$ and $M_2=380$~N\,m$^{-1}$ at $\sigma=0.1$~\AA{}, averaging $S_i^{(12)}$ over two independent planar noise realizations per structure. To ensure that our statistical conclusions hold across the entire generated property space rather than a single configuration, we evaluate 100 generated structures (four structures from each of the 25 target $(M,F_x)$ pairs). We compare the top 10\% most responsive atoms in each structure against its coordination-matched baseline using bond-angle distortion and bond-length range ($r_{\max}-r_{\min}$).
Detailed comparison metric definitions, reference-weighting equation and bootstrap confidence-interval procedures are provided in Supplementary Note~6.




\section{Acknowledgements and funding}\label{sec-109}
This work was supported by the National Natural Science Foundation of China under Grant No. 52541026, No. 52541025. 
We are grateful for computational resources provided by the High Performance Computing Platform of Peking University.






\FloatBarrier

\bibliography{references_20261107}

\end{document}